\documentclass[aps,prd,preprint,tightenlines,superscriptaddress,nofootinbib]{revtex4}
\usepackage{amsmath,amssymb,url}
\usepackage{graphicx,tabularx}

\def\bq{\begin{equation}}
\def\eq{\end{equation}}
\def\ba{\begin{eqnarray}}
\def\ea{\end{eqnarray}}

\newcommand{\etal}{\textit{et al.}}
\newcommand{\ie}{{\em i.e. }}
\newcommand{\eg}{{\em e.g. }}

\newcommand{\cp}[1]{\tilde{\chi}_{#1}^+}
\newcommand{\cm}[1]{\tilde{\chi}_{#1}^-}
\newcommand{\cpm}[1]{\tilde{\chi}_{#1}^\pm}

\newcommand{\wt}{\widetilde}

\newcommand{\po}{\phantom{0}}

\begin{document}

\preprint{}

\date{\today}

\title{Same-Sign Charginos and Majorana Neutralinos at the LHC}

\author{Johan Alwall}
\email{alwall@slac.stanford.edu}
\affiliation{Theoretical Physics, Stanford Linear Accelerator Center,
             Menlo Park, CA, USA}
\author{Dave Rainwater}
\email{rain@pas.rochester.edu}
\affiliation{Dept.~of Physics and Astronomy,
             University of Rochester, Upstate, NY, USA}
\author{Tilman Plehn}
\email{tilman.plehn@cern.ch}
\affiliation{SUPA, School of Physics, University of Edinburgh, Scotland}

\begin{abstract}
We demonstrate the possibility of studying weakly interacting new
particles in weak boson fusion, using the example of supersymmetric
same-sign charginos.  This signal could establish the existence of
Majorana neutralinos and give access to their electroweak couplings.
It can be observed over (supersymmetric) QCD backgrounds provided the
charginos are light and not too close to the squark mass.  We finally
show how same-sign fermion production can be distinguished from
same-sign scalars or vectors arising in other models of new physics.
\end{abstract}

\maketitle


\section{Introduction}

In our search for new phenomena, the LHC is about to start a new era
of testing the incredibly successful and resilient Standard Model. We
know that the Standard Model can only be an effective theory, likely
breaking down around the TeV scale.  One of the possible extensions
which should appear at that scale is supersymmetry (SUSY)~\cite{SUSY}.
Its existence would double the particle spectrum, adding a partner of
opposite spin statistics for every Standard Model particle.

If supersymmetry exists, it must be broken, as we do not see spin
partners of any Standard Model particles~\cite{susy_breaking}.  All
superpartners must therefore be massive compared to their Standard
Model counterparts.  Experiments such as LEP and Tevatron~\cite{run2}
have put stringent bounds on many of the SUSY partner masses.  The LHC
will perform a conclusive search covering masses all the way to the
TeV scale.  In the existing literature we find thorough coverage of
how to conclusively discover SUSY-like signatures at hadron colliders,
primarily via its large production cross sections for the
strongly-interacting squarks and gluino~\cite{Dawson:1983fw,Prospino}.

\bigskip

{\em However}, discovery is only the beginning of LHC physics --- many
alternative scenarios of TeV-scale physics can mimic supersymmetry.
For a long time we have known how to confirm the Majorana nature of
gluinos, provided they are fermions~\cite{SS-dilep}; a similar
strategy for Majorana neutralinos is still missing.  Serious effort
has recently been put into studying how to distinguish between classes
of models, mostly by measuring the masses~\cite{cascades} and
spins~\cite{discriminate} of new particles, mainly in the colored
sector.  Such spectral data can be used to perform TeV-scale model
fits, for example if the spins support a SUSY
hypothesis~\cite{sfitter}.  In comparison, little work has addressed
other quantum number measurements at the LHC.

Typical SUSY spectra show gluino and squarks more massive than the
non-colored superpartners, due to different QCD v. electroweak gauge
coupling evolution from a unification scale, or directly due to the
size of the beta functions of the gauge couplings.  Such heavy
superpartners cascade decay through through successively lighter
superpartners, from colored to colorless, until the cascade terminates
at the lightest supersymmetric particle, or LSP, which is the dark
matter candidate.  A typical (long) squark decay radiates first a
quark to shed its color charge, and then two leptons to finally arrive
at the LSP.  If the gluino is heavier than the squarks, it will have
the same decay chain, plus an extra quark.

\bigskip

While we can accurately measure masses from the decay kinematics in
long cascades, we do not gain any information about the coupling
strengths of the intermediate states, save that they're large enough
to keep the superpartners from being long-lived -- but this is not a
truly useful constraint.  For top quarks, the corresponding issue is
resolved via single-top production, and the analogous process for
stops and sbottoms can establish that the stop-sbottom-$W$ boson
interaction is an electroweak gauge vertex~\cite{Berdine:2005tz}.

Our goal is to observe non-colored superpartner production directly at
LHC, to test those superpartners' electroweak character and study
their quantum numbers independently of cascade decays.  While
Drell-Yan production at the LHC is generally lost in the SM and SUSY
backgrounds, a previous work identified weak-boson-fusion production
as a potentially viable signal~\cite{smadgraph,Datta:2001hv}.  This
channel has been extremely successful in finding ways to study light
Higgs bosons, including for example the size~\cite{duehrssen} and
structure~\cite{higgs_coup} of their couplings.  In the case of
superpartners, the probably most pressing question is the Dirac or
Majorana nature of the neutralino sector, which the weak-boson-fusion
process will allow us to study.


\section{Same-sign charginos}
\label{sec:SSinos}

The production of same-sign charged particles at a hadron collider is
in general a remarkable signal. It requires a balancing of charge in
the final state --- the initial state may have at most charge $\pm 1$
--- which limits it to very few sources.  We will explore each source
for same-sign charginos in turn, starting with the electroweak
production mechanism found in weak boson fusion~\cite{smadgraph}.

In all TeV--scale supersymmetric (MSSM) scenarios, charginos
subsequently decay.  This may be treated as a separate $1\to 2$
on-shell process, which can be included as a branching ratio or as a
fully spin-correlated decay chain (and similarly for any further
decays of the chargino's daughters).  For simplicity, we discuss the
various processes and their associated Feynman diagrams only up to the
produced charginos; their subsequent decay does not alter any of the
production mechanisms or topologies.


\subsection{Weak boson fusion processes}
\label{sub:WBF}
\begin{figure}[ht]
\includegraphics[width=0.25\textwidth]{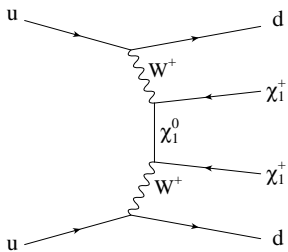}
\caption{Feynman diagram for the pure WBF SUSY process 
$qq^\prime\to qq^\prime\cp1\cp1$ as described in the text.  The
complete set of diagrams sums over all Majorana neutralinos in the
$t$-channel.}
\label{fig:Feyn_WBF}
\end{figure}
In pure weak boson fusion (WBF) a pair of incoming quarks each emit a
weak gauge boson: $W^\pm W^\pm$ for our case of interest.  Because of
the massive gauge boson propagators, the scattered quarks acquire a
transverse momentum typically of the scale of the $W$ mass, $p_T>m_W$.
This is large enough to make the scattered quarks visible as jets in
the detector, albeit at typically small scattering angles, thus far
forward and backward in the detector.  Particles produced in the
fusion process are typically central in the detector, at nearly right
angles to the beam axis, and with similarly high transverse momentum.

Because charginos are fermions, their same-sign production via gauge
boson fusion must be mediated by a $t$-channel neutral Majorana
fermion, to provide the necessary fermion number violation.  In the
MSSM there are four neutralinos.  For each quark-flavor subprocess
there are 8 Feynman diagrams of the topology shown in
Fig.~\ref{fig:Feyn_WBF}.  This set of diagrams is separately gauge
invariant.

\bigskip

WBF same-sign chargino production is most significant for a wino
pair~\cite{smadgraph}, since charged higgsinos have a much smaller
coupling to $W$ bosons.  In most MSSM scenarios in agreement with the
LEP2 limits, the mixing from the SUSY eigenstates to the mass
eigenstates is fairly small.  In that sense the observation of the WBF
signature could establish the gaugino-higgsino nature of the
$\cpm1$--$\cpm2$ hierarchy, an important piece of information in
reconstructing the supersymmetric Lagrangian.


\subsection{Non-WBF electroweak processes}
\label{sub:EW}
\begin{figure}[ht]
\includegraphics[width=0.75\textwidth]{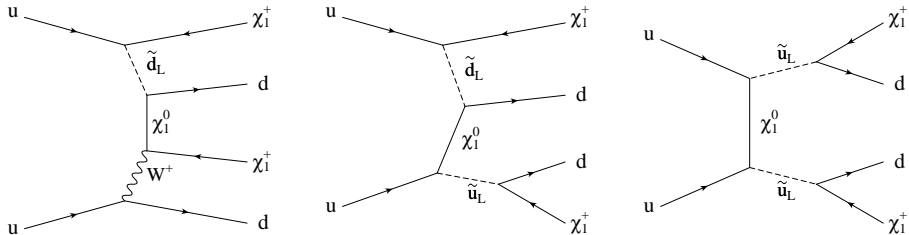}
\vspace{-2mm}
\caption{Representative Feynman diagrams for the electroweak non-WBF 
SUSY process $qq^\prime\to qq^\prime\cp1\cp1$.  The complete set sums
over all Majorana neutralinos in the $t$-channel.}
\label{fig:Feyn_EW}
\end{figure}
The same final state as for WBF processes can occur via electroweak
processes involving non-WBF diagrams, shown by the representative
Feynman diagrams of Fig.~\ref{fig:Feyn_EW}.  We observe non-resonant
$t$-channel diagrams, singly-resonant squark and doubly-resonant
squark processes.  The latter numerically dominate, but to properly
account for off-shell effects while maintaining gauge invariance we
perform a complete calculation.  This completeness will become
important once we impose kinematic cuts to suppress on-shell squarks.


\subsection{QCD processes}
\label{sub:QCD}
\begin{figure}[ht]
\includegraphics[width=0.75\textwidth]{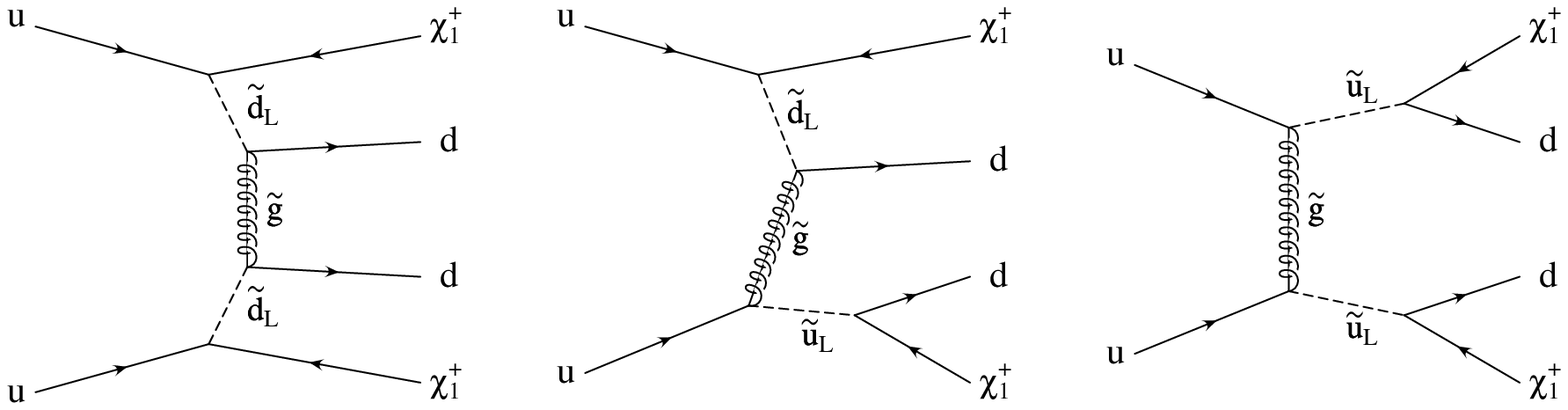}
\vspace{-2mm}
\caption{Representative Feynman diagrams for the QCD SUSY process 
$qq^\prime\to qq^\prime\cp1\cp1$.}
\label{fig:Feyn_QCD}
\end{figure}
The dominant background before any kinematic cuts arises from cascade
decays of heavy colored squarks (and gluinos, if heavier than
squarks), as discussed in the introduction; see
Fig.~\ref{fig:Feyn_QCD}.  For example, LHC will provide an enormous
flux of pairs of valence $u$ quarks, which can scatter to a pair of
same-sign up squarks via a $t$-channel Majorana
gluino.\footnote{Same-sign charginos (or same-sign leptons, if
charginos decay promptly) from QCD processes can be taken as evidence
of the Majorana nature of the gluino~\cite{SS-dilep}, once its
fermionic spin character is established~\cite{discriminate}.}  Gluino
pairs may also decay to same-sign squarks, giving the same final state
modulo extra jets; likewise for squark--gluino mixed production.  The
different processes might be distinguished using the jet
multiplicity~\cite{danish}.  All QCD processes occur at huge rates
compared to both electroweak sources of like-sign charginos, despite
the higher final-state masses and consequent phase space suppression.
As in the electroweak non-WBF case, the doubly-resonant component
dominates, but we include all possible QCD amplitudes to correctly
account for off-shell effects.


\section{Signal and backgrounds}
\label{sec:SvB}
\begin{table}[t]
\begin{tabular}{|c||>{\po}l<{\po}|>{\po}l<{\po}|>{\po}l<{\po}|>{\po}l<{\po}
                   |>{\po}l<{\po}|>{\po}l<{\po}|>{\po}l<{\po}|>{\po}l<{\po}
                   |>{\po}l<{\po}|>{\po}l<{\po}|}
\hline
\, SPS \, & 1a & 1b & 2 & 3 & 4 & 5 & 6 & 7 & 8 & 9 \\
\hline
$\chi^+_1\chi^+_1$ 
& 0.93 & 0.22  & 0.48 & 0.23  & 0.51 & 0.57 & 0.067 & 0.077 & 0.31  & 0.88 \\
\hline
$\chi^-_1\chi^-_1$ 
& 0.28 & 0.056 & 0.13 & 0.058 & 0.14 & 0.16 & 0.017 & 0.020 & 0.083 & 0.25 \\
\hline
\end{tabular}\centering
\caption{Cross sections~[fb] for WBF opposite-sign and same-sign 
chargino pair production at LHC, for all MSSM benchmark SPS points,
without cuts, from Ref.~\protect\cite{smadgraph}.  Cross sections are
shown to two significant digits.}
\label{tab:SPS_xx}
\end{table}
We begin by reviewing the WBF same-sign chargino cross sections
calculated in Ref.~\cite{smadgraph}; the results are repeated in
Table~\ref{tab:SPS_xx} for convenience.  With the exception of a few
SPS points~\cite{sps}, the cross sections are comparable, of order
1~fb and falling mostly in a range of a factor of three of each other.
These leading-order total cross sections are without cuts.
Observation at LHC would depend on the rate for a given final state,
which would typically require leptons for charge identification.  If
the chargino decays to lepton plus slepton, this could be done with
high efficiency (near $100\%$ if electron or muon, less for tau).  If
instead it decays to $W$ boson plus neutralino (typically the lightest
one, the LSP), there would be a larger hit in signal rate due to the
requirement to observe leptonic $W$ decays.

In some MSSM scenarios, notably anomaly-mediated supersymmetry
breaking, the chargino is long-lived due to a near-degeneracy with the
lightest neutralino, the LSP.  Long-lived charged massive particles,
or CHAMPs, are searched for at the Tevatron and would be readily
observable~\cite{Fairbairn:2006gg} in the LHC detectors,
ATLAS~\cite{ATLAS} and CMS~\cite{CMS}.  Heavy gluinos from gluon
fusion are produced very close to threshold, so one has to require
$\beta\gtrsim0.6-0.8$. In the case of WBF charginos this captures the
bulk of the signal for charginos light enough to be produced at a
sufficient rate, so we do not explicitly impose such a cut.

\bigskip

Our stable-chargino scenario provides the ``best'' possible signal.
First, because of the very high efficiency to capture such events in
data.  Second, because cascade decay chains are fully reconstructible,
so all superpartner masses would be known.  We can impose tailored
invariant-mass cuts to remove the squark and gluino backgrounds.  We
examine this scenario first as a baseline to all others, but will find
that backgrounds are generally not a problem even when the chargino
cannot be reconstructed.  In general, we will not focus on the
kinematics of the centrally produced charginos --- instead, we follow
the spirit of a similar Higgs-coupling analysis~\cite{higgs_coup} and
rely on the tagging-jet kinematics to analyze the events.

Specifically, we investigate the benchmark point SPS9~\cite{sps}, an
anomaly-mediated supersymmetry-breaking scenario with naturally
long-lived wino-like charginos $\cpm1$.  The lighter chargino has a
mass of 197.4~GeV and the light-flavor squark masses are 1.3~TeV, with
a lighter gluino.  Gluino production is not a background, because
gluinos decay to either a top quark plus stop or bottom quark plus
sbottom with a radically different final state which can be easily
vetoed.  We emphasize that SPS9 is only a toy example to demonstrate
the utility of this signature with a minimum of complication.  Our
analysis is equally valid in any other scenario, provided the
production rate for same-sign charginos (or their decay products after
branching ratios) is sufficient for observation at high luminosity.

\bigskip

For all our calculations we use the event generator {\sc
madevent4}~\cite{madevent} with its MSSM extension~\cite{smadgraph}.
We consistently utilize the leading-order parton densities
CTEQ6L1~\cite{Pumplin:2002vw}.  For all electroweak processes we
select the minimum transverse momentum of the tagging jets as the
factorization scale, $\mu_F={\rm min}(p_T(j))$.  For QCD processes we
use the squark mass for the factorization and renormalization scales,
$\mu_F=\mu_R=m_{\wt{q}}$, as suggested by NLO
calculations~\cite{Prospino}.

In addition to an assumed $b$ jet veto to remove (supersymmetric)
heavy-flavor backgrounds, we apply the usual weak-boson-fusion cuts
for the tagging jets.  On top of those we require minimal cuts for the
charginos to satisfy detector requirements for observability and
tracking.  We expect these chargino cuts to have a similar effect as
cuts on possible chargino decay products.  None of our later results
depend in any way on the chargino cuts.  The basic level cuts consist
of minimum transverse momentum and maximum absolute rapidity:
\bq\label{eq:cut-A}
p_T(j) > 20 \, {\rm GeV}, \quad |\eta(j)| < 4.5, \quad
p_T(\cpm1) > 10 \, {\rm GeV}, \quad |\eta(\cpm1)| < 2.5 \; .
\eq
To make use of the inherent characteristics of WBF particle
production, namely forward-scattered quark jets with large rapidity
separation between them, and central production of the electroweak
objects, we impose a jet separation cut and require the colorless
objects to lie between the jets~\cite{Asai:2004ws}:
\bq\label{eq:cut-B}
|\eta(j_1)-\eta(j_2)| > 3.0 \, , \quad
\eta(j)_{\rm min} < \eta_{\cpm1} < \eta(j)_{\rm max} \; .
\eq
Additionally, we impose an invariant mass cut on all combinations of
one jet with one chargino around the known squark mass, which may
easily be done for long-lived massive charged
particles~\cite{Fairbairn:2006gg}.  We study two more-or-less
aggressive versions of this cut, which almost completely removes the
QCD and electroweak squark-production backgrounds:
\bq\label{eq:cut-C}
|M(j,\cpm1)-m_{\wt{u}}| > 30 (50) \, {\rm GeV} \, .
\eq
This cut of course assumes long-lived charged particles.  We give
results for both options, as well as only WBF cuts, to show how cut
optimization may affect signal rate and the signal-to-background
ratio, $S/B$.


\subsection{Cross sections}
\label{sub:LL}
\begin{table}[t]
\begin{center}
\begin{tabular}{|l|c|c|c|}
\hline
cuts
& WBF cuts
& $|m_{j\chi}-m_{\wt{q}}|>30$~GeV 
& $|m_{j\chi}-m_{\wt{q}}|>50$~GeV 
\\
\hline
\hline
All EW     & 1.138 (0.286) fb & 0.847 (0.226) fb & 0.786 (0.213) fb \\
\hline
WBF        & 0.825 (0.220) fb & 0.766 (0.206) fb & 0.724 (0.197) fb \\
\hline
EW non-WBF & 0.261 (0.053) fb & 41.4  (8.52)  ab &  23.1 (4.76)  ab \\
\hline
QCD        & 0.259 (0.040) fb &  8.70 (1.58)  ab &  3.66 (0.775) ab \\
\hline
$S/B$      & 1.6/1 (2.4/1)    & 15/1  (20/1)     &  27/1 (36/1)     \\
\hline
\end{tabular}
\end{center}
\caption{LHC cross sections for the WBF same-sign chargino signal 
$\cp1\cp1jj$ ($\cm1\cm1jj$), electroweak and QCD backgrounds at SPS9, for
various levels of kinematic cuts described in the text.  We also show
the signal-to-background ratio $S/B$.}
\label{tab:xsecs}
\end{table}
Our cross section results for $\cp1\cp1jj$ and $\cm1\cm1jj$ production
at LHC with various levels of kinematic cuts for SPS9 are shown in
Table~\ref{tab:xsecs}.  The QCD and electroweak non-WBF backgrounds
are each slightly less than half the size of the signal already after
basic WBF cuts. The excellent ratio $S/B$ is promising, even when the
number of signal events is small.  In a long-lived chargino scenario
the event may be completely reconstructed, allowing the imposition of
an invariant mass cut to remove the squark poles.  If such a rejection
cut is possible, the backgrounds become truly negligible.

If the charginos decay, this cut would likely turn into something like
a transverse-mass cut with lower efficiency.  Moreover, there would be
some efficiency loss in selecting leptonic final states, but standard
techniques in WBF~\cite{Asai:2004ws} would provide for further
significant reduction of the backgrounds.  Such generalization is
however beyond the scope of this first paper.

\bigskip

Given that the efficiency to tag two forward jets as well as the two
central charged tracks is collectively about
$60\%$~\cite{Asai:2004ws}, our signal requires the full LHC luminosity
of 300~fb$^{-1}$, especially to obtain good statistics in the
kinematic distributions.  WBF production of exotic particles is a
natural case for the high-luminosity environment of the
SLHC~\cite{Gianotti:2002xx}, when parameter studies will become more
and more the focus for the experiments.  However, forward-jet tagging
at those luminosities is not yet fully understood, so we limit
ourselves to the LHC design luminosity before its planned upgrade.
There, we expect a few hundred signal events with high purity --- more
than enough to perform ``precision'' measurements in kinematic
distributions, given negligible backgrounds.  The rate uncertainties
would be around ${\cal O}(5\%)$ statistically, probably with similar
systematic uncertainties.  Parton-density and higher-order QCD
uncertainties are known to be of that size or smaller from WBF Higgs
and vector boson studies~\cite{WBF-NLO}.


\subsection{Kinematic distributions}
\label{sub:SvB-kin}

\begin{figure}[t]
\begin{center}
\includegraphics[scale=0.4]{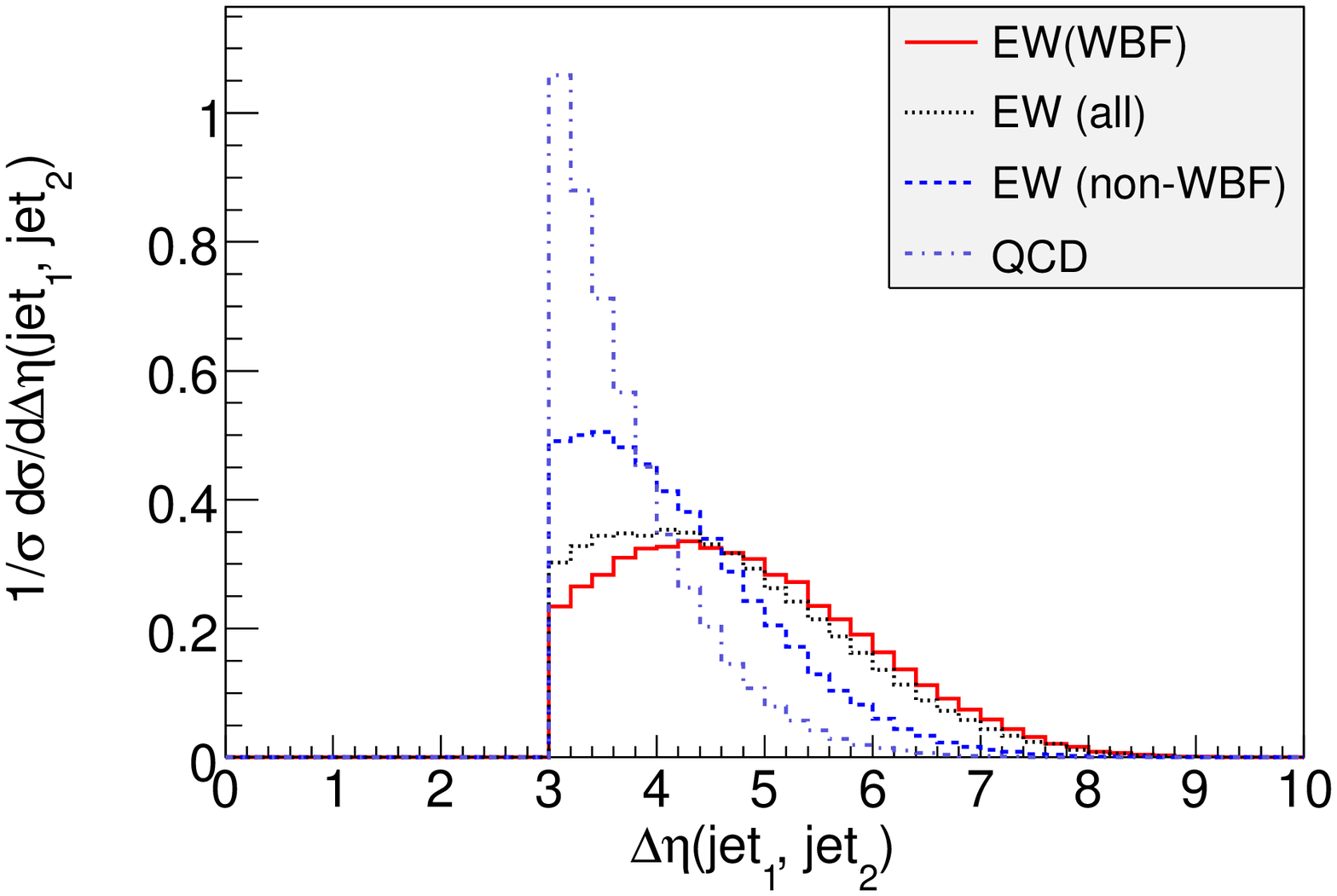}
\includegraphics[scale=0.4]{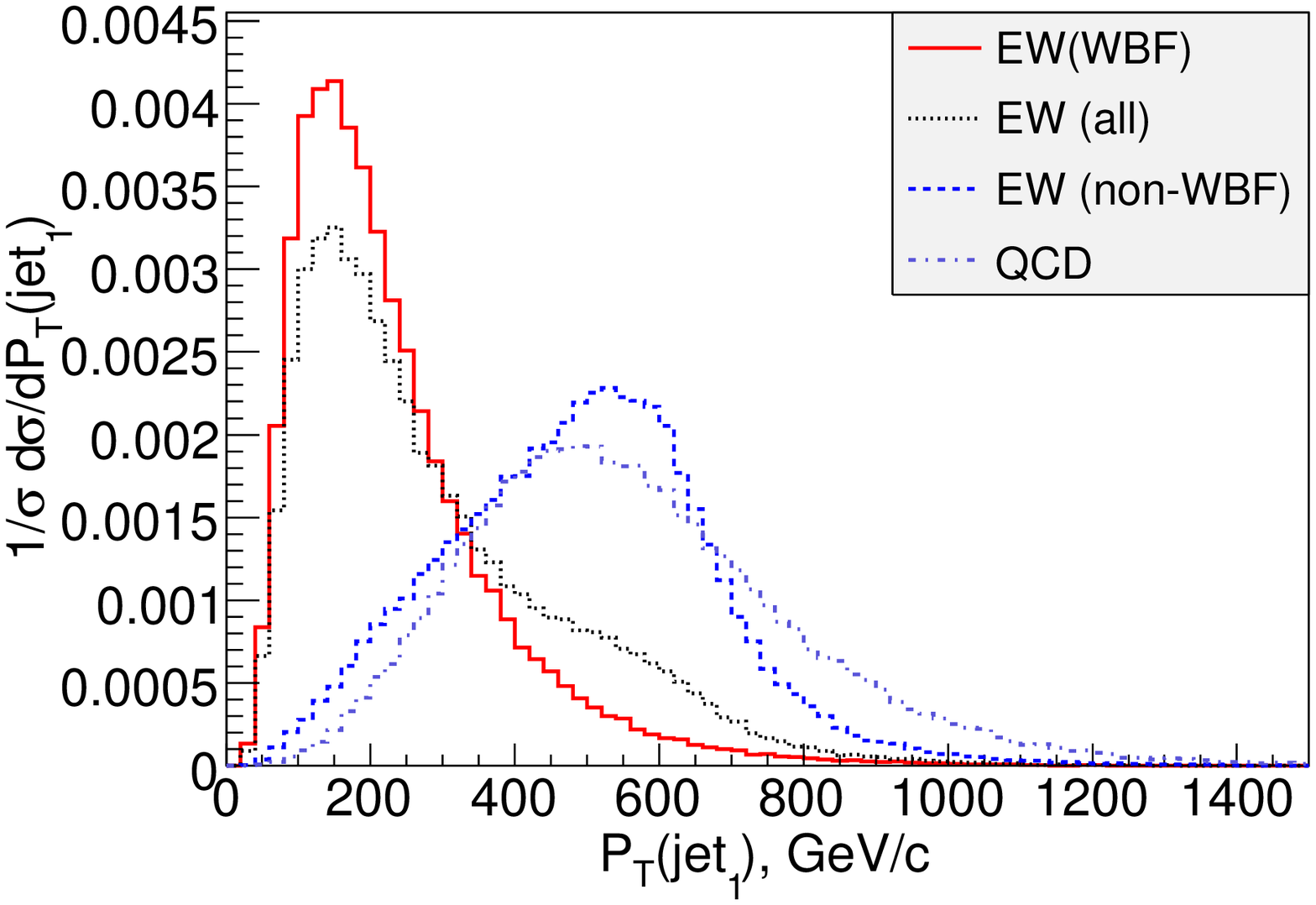}
\includegraphics[scale=0.4]{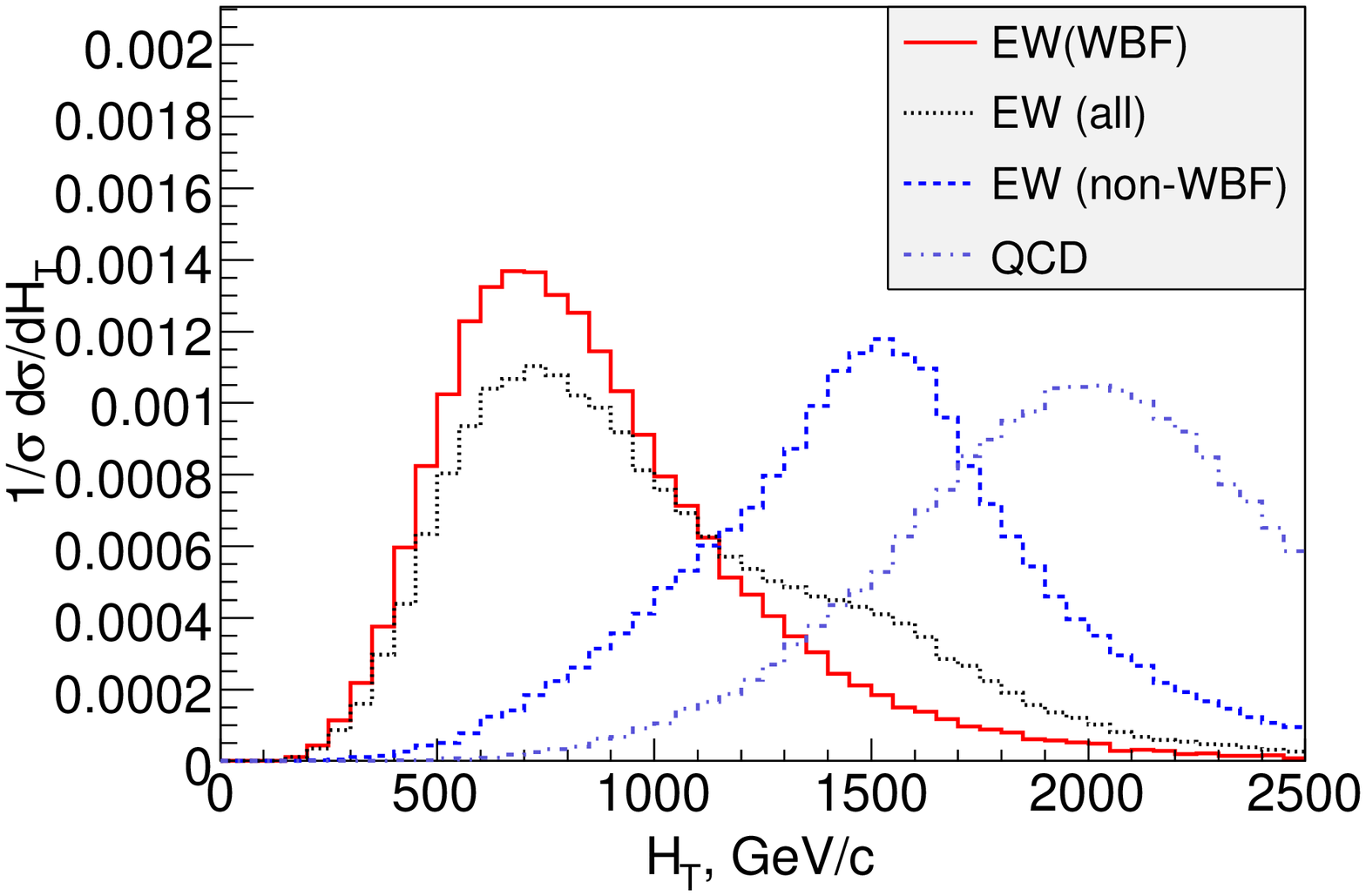}
\includegraphics[scale=0.4]{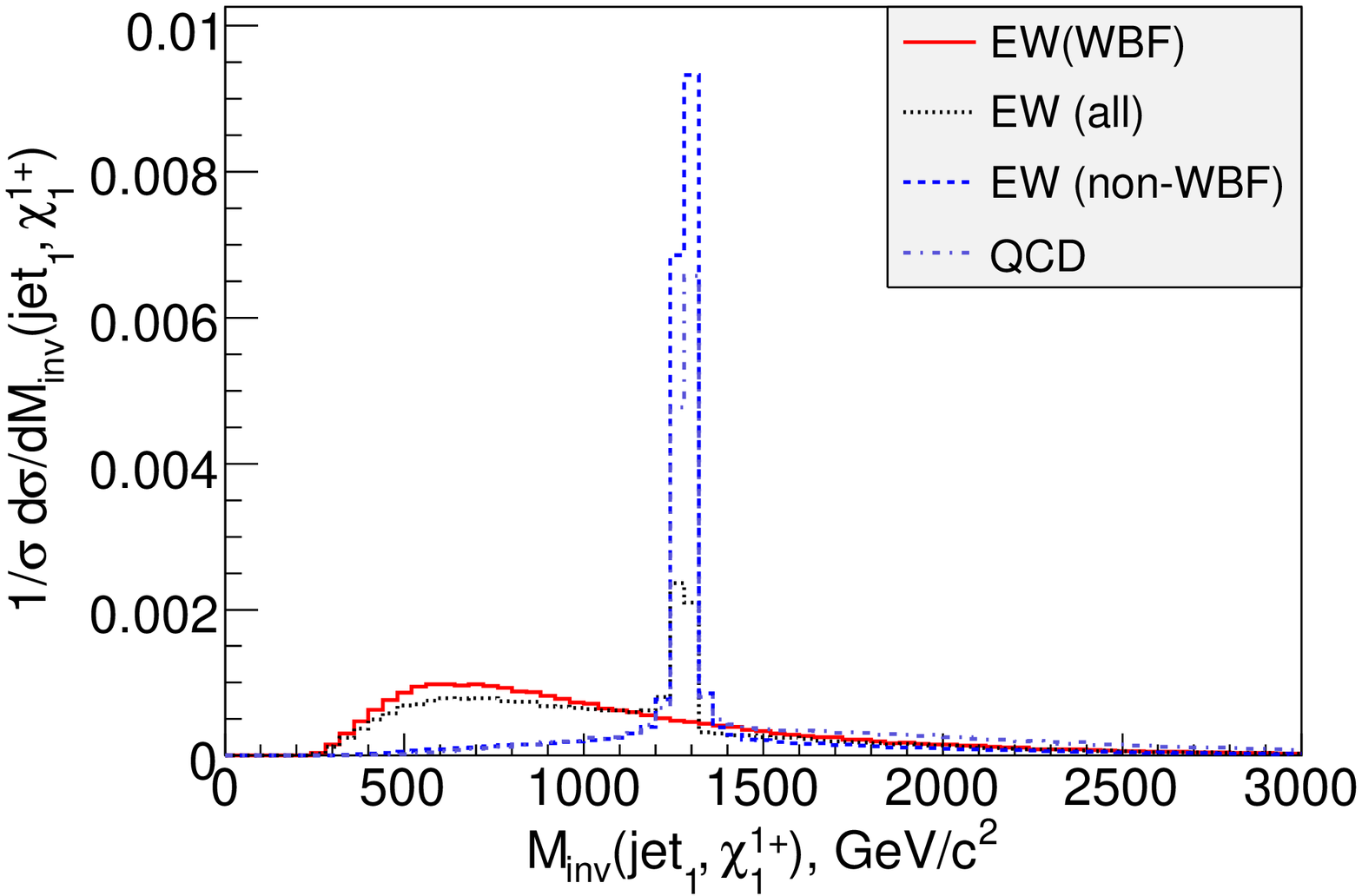}
\end{center}
\vspace{-6mm}
\caption{Distributions for the WBF same-sign chargino signal at SPS9, 
including the electroweak and QCD backgrounds from squark production.
The jets are ordered according to their transverse momentum.  Only WBF
cuts Eqs.~(\protect\ref{eq:cut-A},\protect\ref{eq:cut-B}) are used.
The invariant mass distribution can be used for long-lived charginos
only.  All distributions are normalized to unity; see
Table~\protect\ref{tab:xsecs} for total rates.}
\label{fig:SvB}
\end{figure}

We show two useful kinematic distributions of the final-state forward
tagging jets in Fig.~\ref{fig:SvB}, as well as the total deposited
transverse energy of all observed objects, $H_T=\mathop\sum_i
E_{T_i}$, and the jet--chargino invariant mass.  For these plots, we
impose only the WBF cuts of Eqs.~(\ref{eq:cut-A},\ref{eq:cut-B}).  All
curves are normalized to unity to emphasize the gross distinguishing
characteristics of WBF v. non-WBF electroweak and QCD production.  As
noted earlier, the backgrounds arise primarily from heavy squark
decays, so they give much harder jets (and charginos) for a typical
mass separation.  This is also the reason for the backgrounds' far
larger $H_T$ compared to the WBF continuum signal. The leading
chargino transverse momentum distribution is almost identical to that
for the leading tagging jet.  Other standard WBF distributions, such
as the azimuthal angle between the tagging jets, $\triangle\phi_{jj}$,
are only marginally discriminating.

\bigskip

In a long-lived chargino scenario, all these distributions may be used
to suppress the backgrounds to a truly negligible level relative to
the signal, as we found in Table~\ref{tab:xsecs}.  If the chargino
instead decayed promptly, then all curves in $H_T$ would shift to the
left by some likely universal amount, to account for the unobserved
LSP pair.  The point is that while all of these distributions would
change, the shifts would be very similar for signal and backgrounds,
thus retaining the same basic distinctions and separating power.  Only
the invariant mass of a chargino plus the leading jet is valid
exclusively for long-lived scenarios.  But, as can be seen from the
other distributions, obtaining even better $S/B$ ratios achieved with
the first level of standard WBF cuts would be straightforward, also
with decays.

We thus do not anticipate any serious background issues, at least in
scenarios where the squarks are appreciably heavier than the
charginos.  For many SUSY models, $S/B$ will be of order 1/1 from the
WBF cuts alone, Eqs.~(\ref{eq:cut-A}--\ref{eq:cut-B}).  Very little
effort would be needed to enhance the signal v. background separation
further.


\section{Discriminating between new-physics models}
\label{sec:discrim}

Making new physics discoveries at the LHC immediately means facing the
arduous task of determining what we actually see.  If we observe
same-sign charged particles in weak boson fusion we cannot simply
assume that they are supersymmetric charginos; alternative hypotheses
must be tested to reach a meaningful conclusion. We already know that
other objects can at first glance appear to be gluinos or
squarks~\cite{UED-fake}.  Even if heavy colored particle decays were
determined to be of the right spin for
supersymmetry~\cite{discriminate}, it can be hard to determine
chargino and neutralino candidate spins in cascade decays, not to
mention the Majorana or Dirac nature of such weakly interacting new
fermions.  Our goal is to show that kinematic distributions can be
used to discriminate between fermionic same-sign particle production
in WBF, and scalars or vectors.  (We ignore higher-spin states in good
taste.)

\medskip

To formulate our stable-scalar hypothesis, we use the MSSM two-Higgs
doublet model, as implemented in {\sc
madevent4}~\cite{madevent,smadgraph}.  A general two-Higgs doublet
model (also implemented) could be used as well, but this does not
change the spin structure, and in any case we consider only normalized
distributions.  For the spin hypothesis comparison we assume the
charged Higgs to be stable on detector timescales due to a near mass
degeneracy with its decay products; it may also decay promptly, and
all kinematic distributions alter in a way similar to that described
for the fermionic chargino case in Sec.~\ref{sub:SvB-kin}.

For the vector case, we implement a generic model with a neutral
$Z^\prime$ and a charged $W^\prime$ pair.  If done rigorously, this is
not entirely straightforward, as we would need to begin with a larger
gauge group, such as $SU(2)_L\times SU(2)_R$, and break it to
$SU(2)_L$.  We would also have to be careful about any additional
matter content, which may be necessary depending on the underlying
group structure and the breaking mechanism~\cite{little_higgs}.  Since
we normalize the scalar and vector cross sections to the fermionic
chargino rate and analyze exclusively normalized distributions, we do
not worry about such details.  Instead, we use a toy model based on an
additional Little-Higgs-type gauge sector $W^\prime/Z^\prime$ with $T$
parity, which makes the $W^\prime$ stable on detector timescales.  To
preserve unitarity at high energies, we include a $T$-odd scalar
$H^\prime$ and $T$-odd heavy quarks $u^\prime,d^\prime$, etc.  The
Feynman rules are the same as for the corresponding Standard Model
vertices in all cases.  With our $H^\prime$ we verify unitarity
conservation in the process $W^+W^+\to W^{\prime+}W^{\prime+}$.  At
the LHC, we find that for unitary WBF $W^{\prime+}W^{\prime+}jj$
production the $T$-odd Higgs is not necessary; removing it does not
yield a noticeable change in results, because the parton densities
restrict the cross section at energies where the Higgs exchange
becomes important.  However, fermionic partners {\it must} be present
for our coupling structure.  They provide the gauge cancellations
necessary for unitarity at energies well below a strong dynamics
scale, as will be discussed below.  Note that not all Little Higgs
models contain all these states; instead, strong dynamics is expected
to appear at the few-TeV scale.

\bigskip

To limit our analysis to actual spin effects, we set all final-state
same-sign charged particles masses to the chargino mass of SPS9
(197~GeV).  Moreover, we normalize all rates, as is common in similar
LHC spin studies~\cite{discriminate}.  We recognize that in general
new-physics scenarios, charged scalars and charged vectors are
unlikely to be long-lived.  However, as stressed above, our analysis
in no way relies on this assumption. To make this obvious we show
distinguishing kinematic distributions only for the tagging jets.  It
turns out that they alone can clearly discriminate between the various
spins.

\begin{figure}[t]
\begin{center}
\includegraphics[scale=0.4]{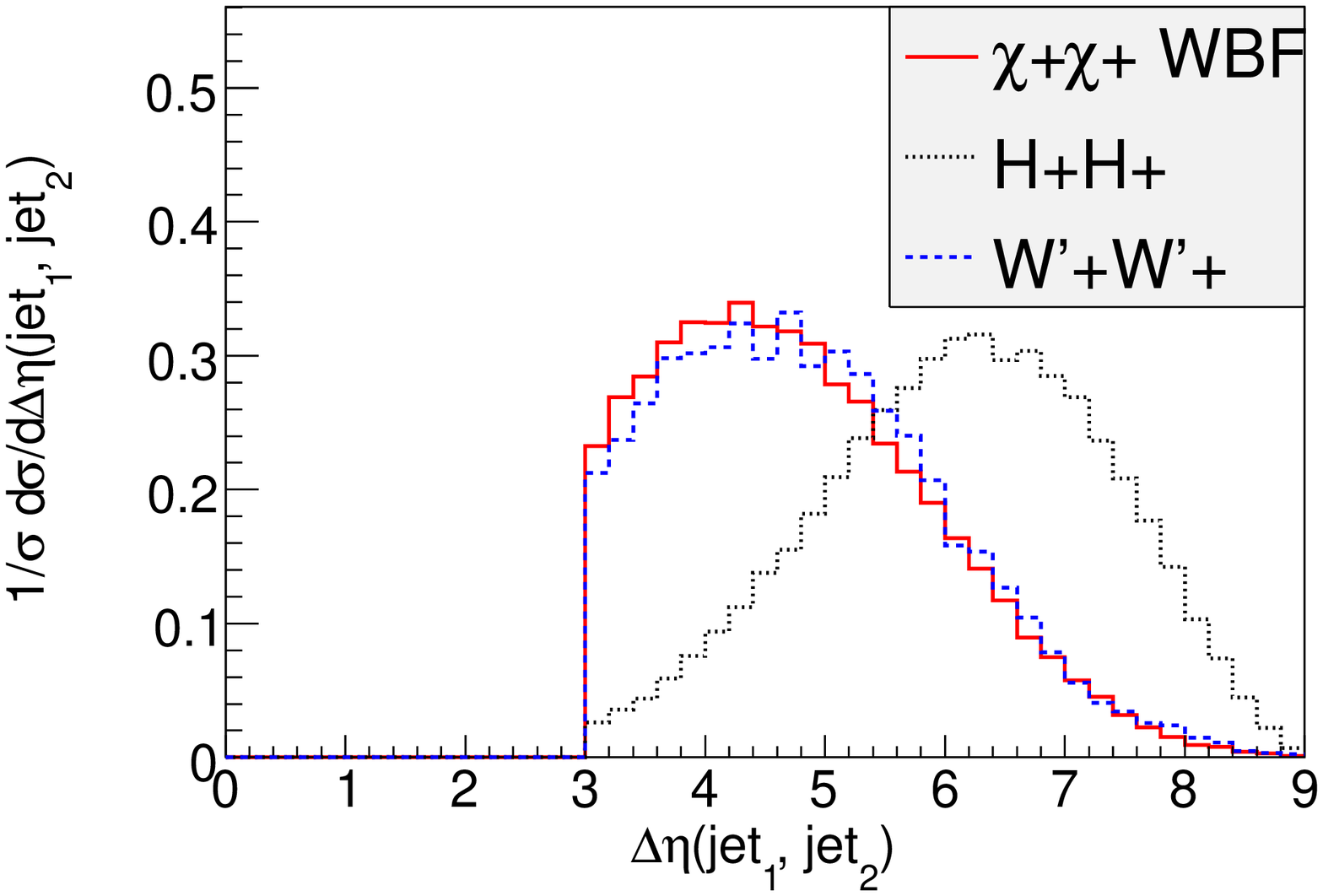}
\includegraphics[scale=0.4]{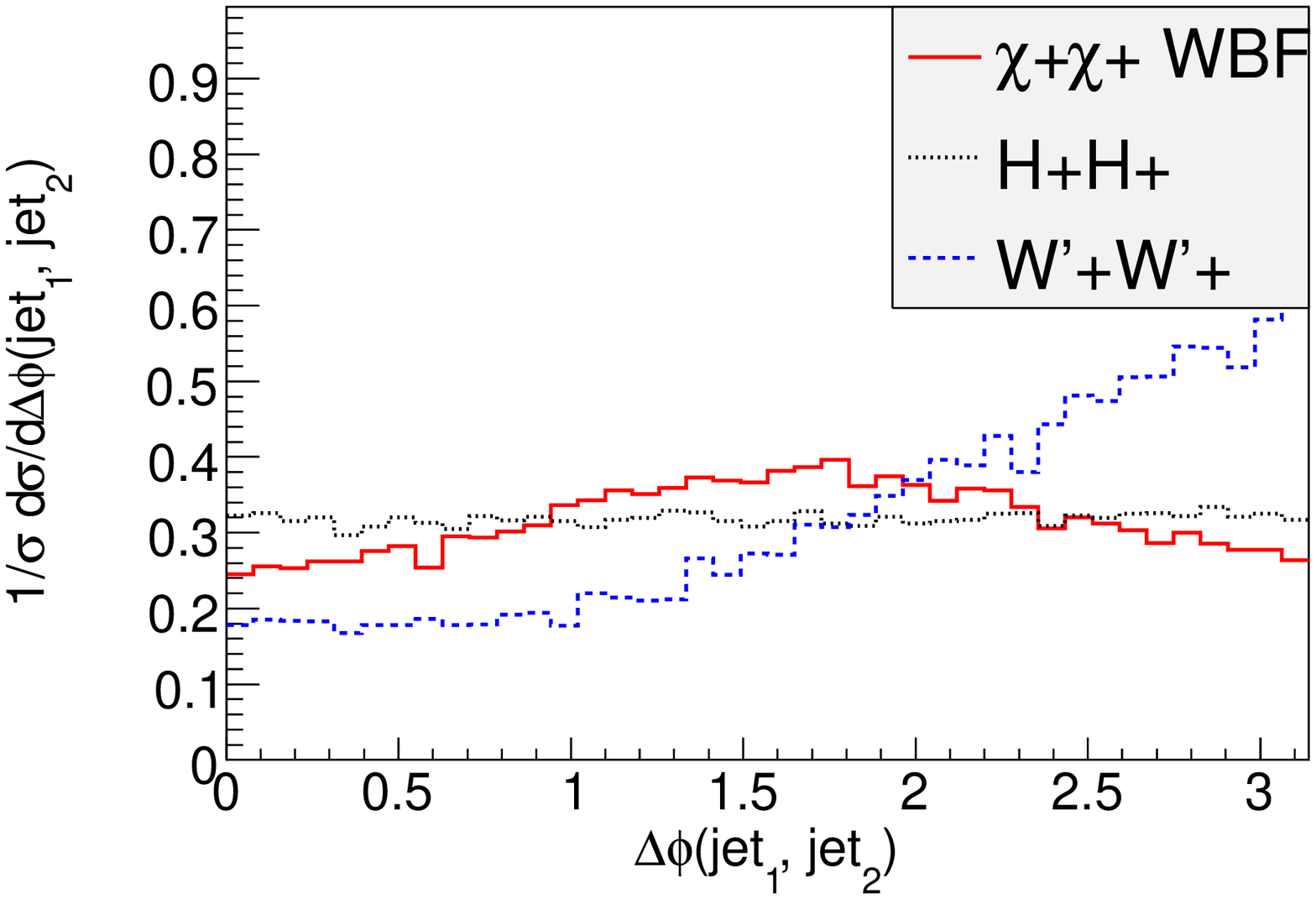}
\includegraphics[scale=0.4]{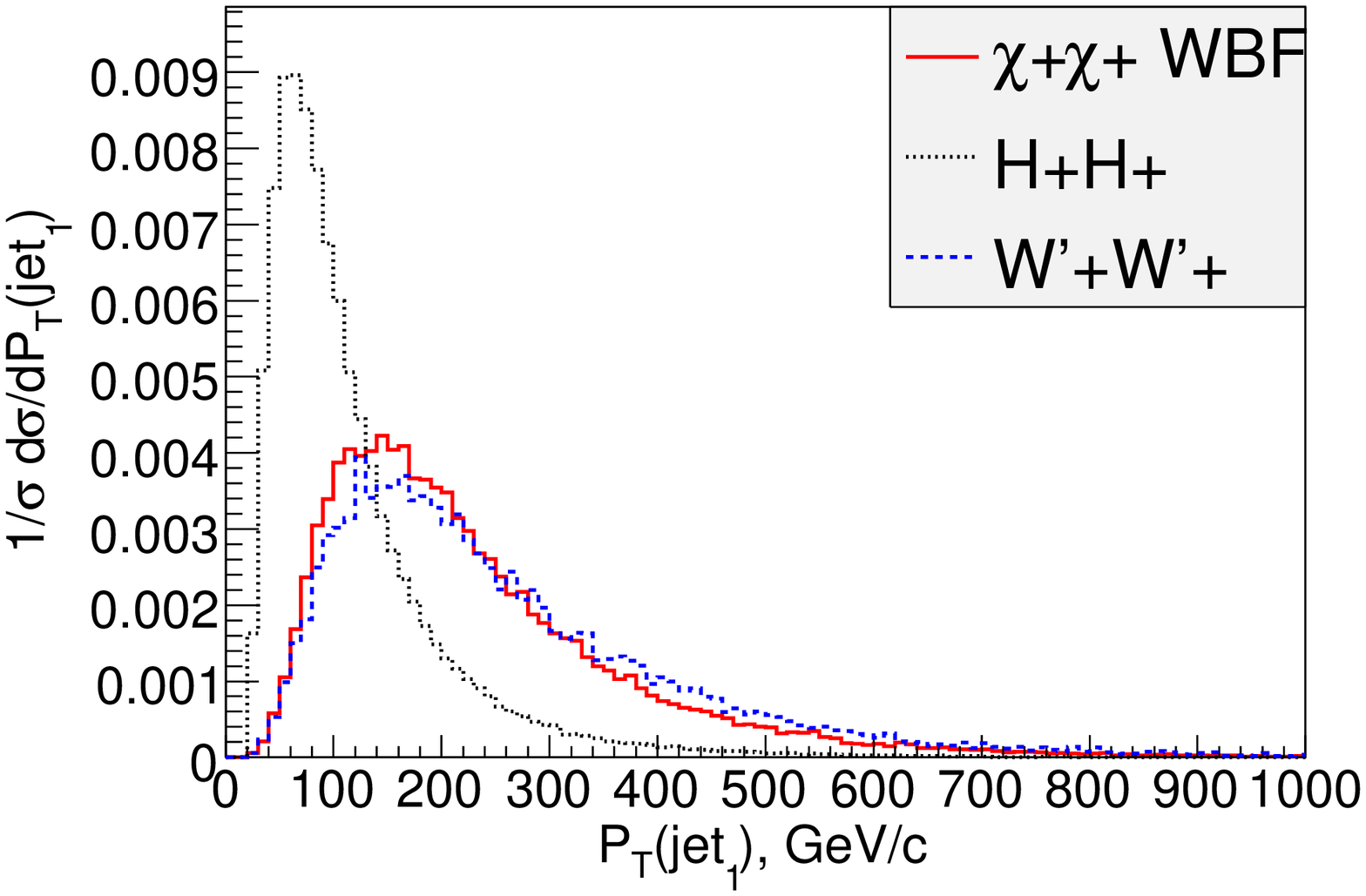}
\includegraphics[scale=0.4]{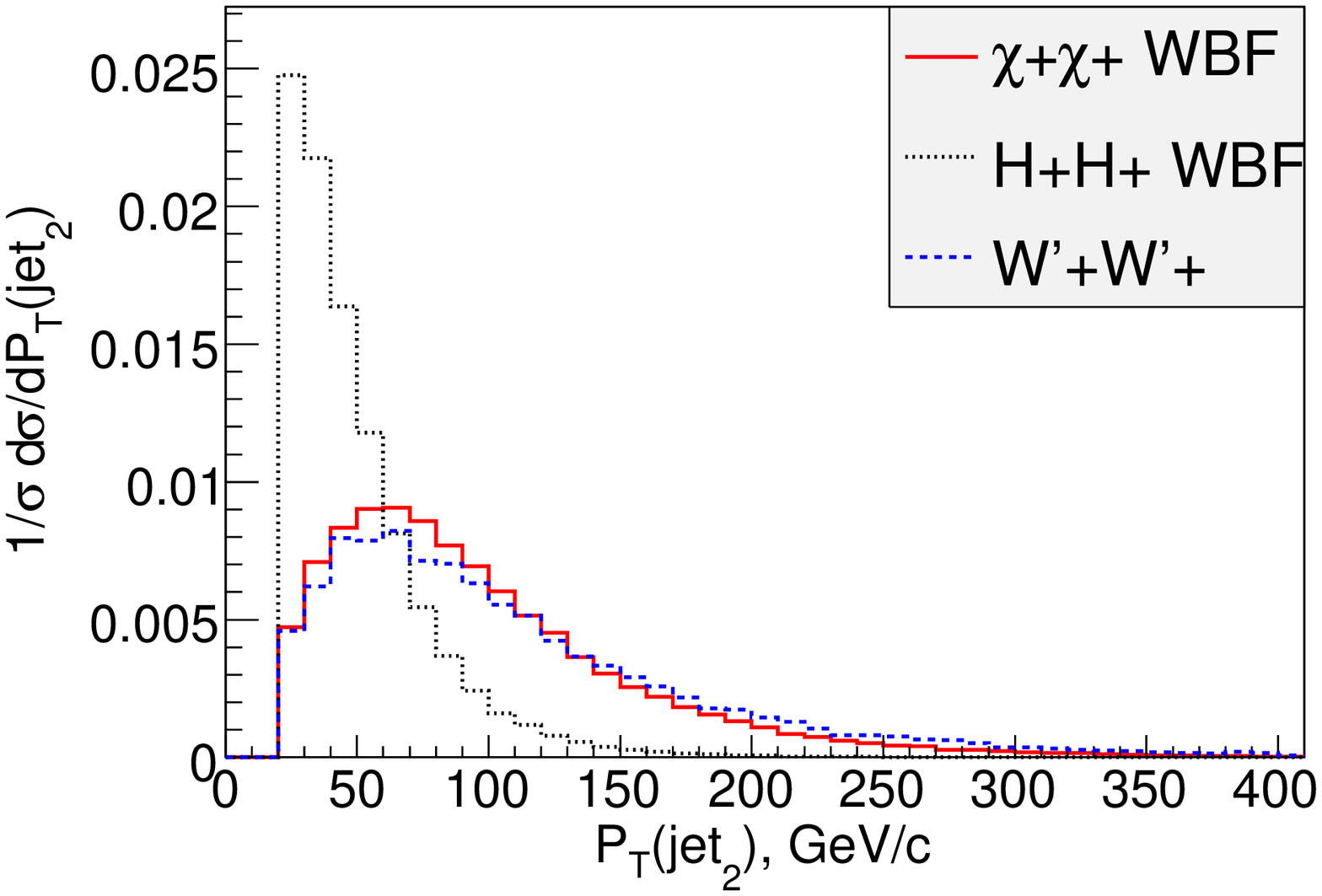}
\end{center}
\vspace{-5mm}
\caption{Kinematic distributions for the tagging jets in WBF production
of charginos, scalars and vectors.  None of the distributions rely on
the charged particles being long-lived.}
\label{fig:comp-jet}
\end{figure}

Fig.~\ref{fig:comp-jet} shows four distributions for the two forward
tagging jets: two angular correlations and two transverse momenta.
All of them are independent of the long-lived nature of the charged
particles.  We first notice that the scalar case is markedly different
from either the fermion or vector cases in all distributions.  This
arises from the virtual $W$ boson emitted from the incoming quarks.
The scalar sector couples to the longitudinal (Goldstone) mode of the
virtual $W$ boson, which has a distinct preference for small-angle
emission, {\it i.e.} a more forward, low-$p_T$ tagging jet.  Fermions
and vectors have no such preference, so the transverse modes
contribute much more prominently.  If we consider a quark with energy
$E$ radiating a vector boson with energy $xE$ and transverse momentum
$p_T$, the probability of collinear radiation of a transverse or
longitudinal $W$ boson can be approximated by~\cite{tao}:
\begin{alignat}{5}
P_T(x,p_T) &\sim \frac{g_V^2 + g_A^2}{8 \pi^2} \;
                 \frac{1+(1-x)^2}{x} \;
                 \frac{p_T^2}{(p_T^2 + (1-x) \, m_W^2)^2} 
          &&\longrightarrow  
                 \frac{g_V^2 + g_A^2}{4 \pi^2} \;
                 \frac{1+(1-x)^2}{2x} \;
                 \frac{1}{p_T^2} 
                 \notag \\
P_L(x,p_T) &\sim \frac{g_V^2 + g_A^2}{4 \pi^2} \;
                 \frac{(1-x)^2}{x} \;
                 \frac{m_W^2}{(p_T^2 + (1-x) \, m_W^2)^2} 
          &&\longrightarrow  
                 \frac{g_V^2 + g_A^2}{4 \pi^2} \;
                 \frac{(1-x)^2}{x} \;
                 \frac{m_W^2}{p_T^4} 
\end{alignat}
The couplings $g_{A,V}$ describe the gauge coupling of the $W$ bosons
to the incoming quarks.  The last approximation assumes large
transverse momentum $p_T\gg (1-x)m_W$, describing the upper end of the
$p_T$ spectrum.  In this limit the radiation of longitudinal $W$
bosons falls off sharper than the radiation of transverse $W$ bosons,
\ie, the tagging jets associated with Higgs production are softer than
the tagging jets associated with fermionic charginos or with vectors.
This is confirmed in the first, third and fourth panels.  Note that in
the more realistic of the $x$ limits, $x\ll 1$, where the tagging jets
carry most of the energy, there is no difference in the $x$ behavior
of the two spectra.  Unfortunately, none of these three distributions
distinguishes between objects which couple to the transverse modes,
{\it i.e.}  between fermions and vectors, which can be understood from
this simple approximation.

\begin{figure}[ht!]
\begin{center}
\includegraphics[scale=0.4]{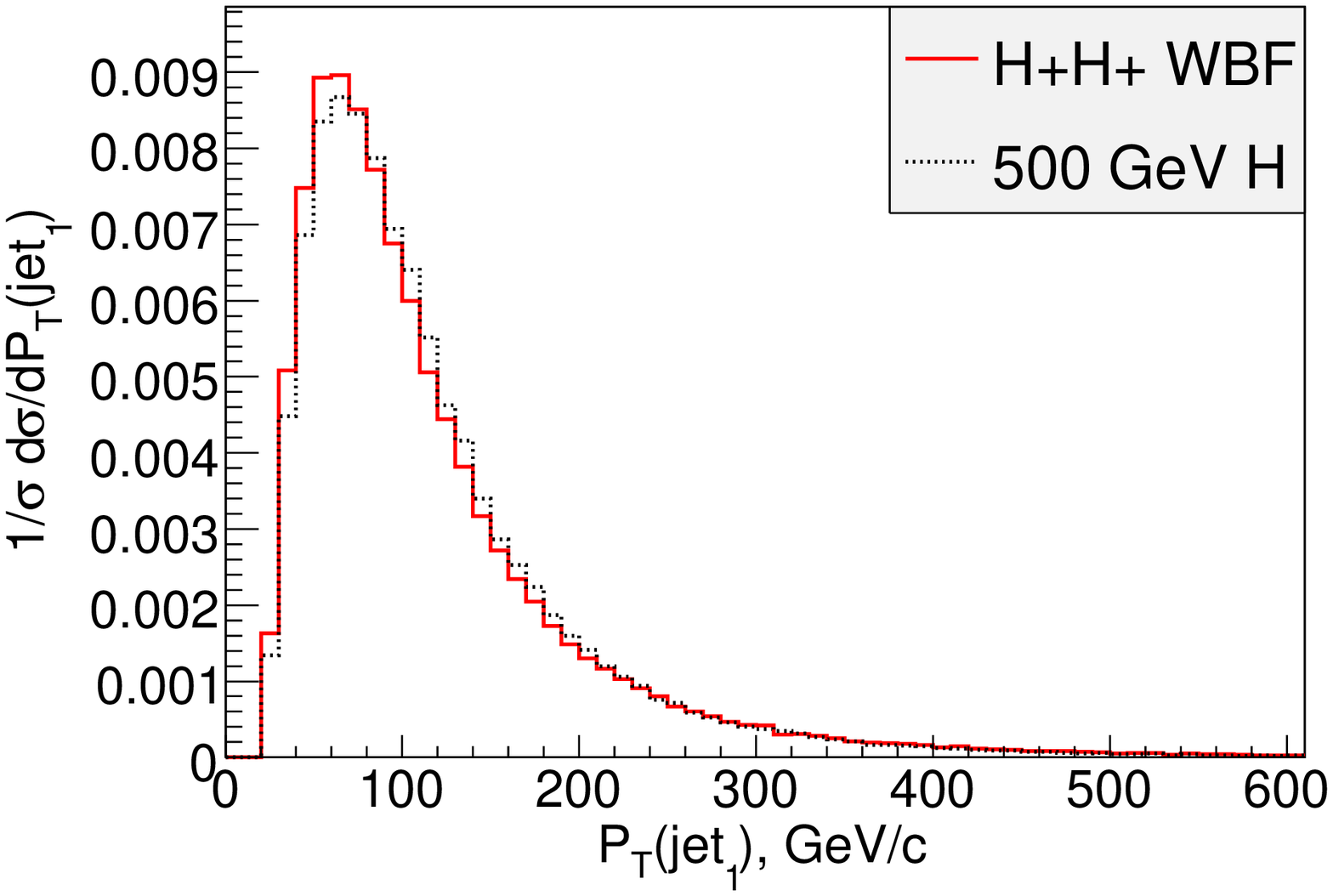}
\includegraphics[scale=0.4]{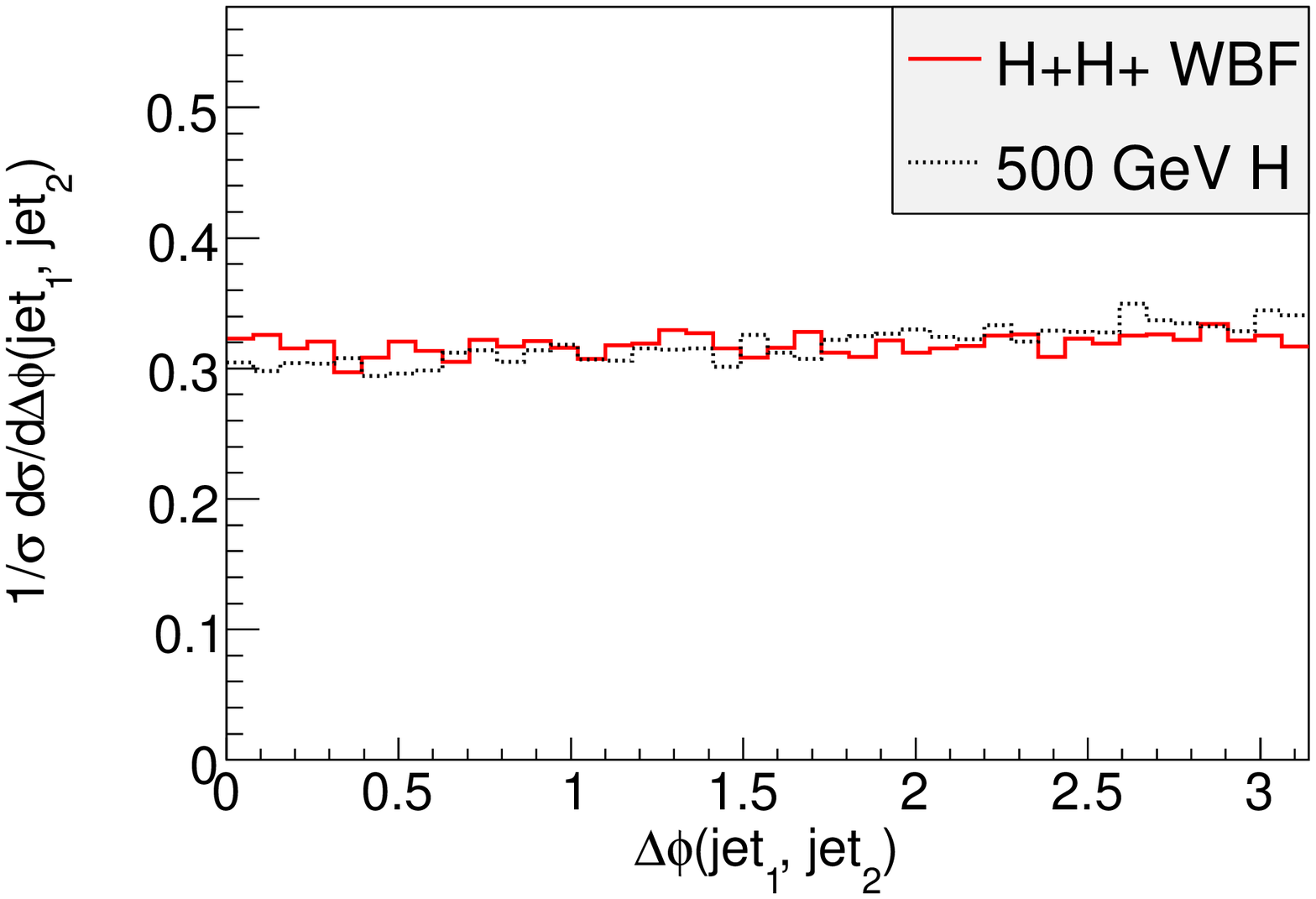}
\end{center}
\vspace{-5mm}
\caption{Distributions comparing charged scalar pair production with a
single neutral scalar of 500~GeV, to demonstrate identical spin
structure.}
\label{fig:scalar}
\end{figure}

The second panel of Fig.~\ref{fig:comp-jet} saves the day.  It shows
the azimuthal angular separation, which has a slight enhancement for
charginos at $\triangle\phi_{jj}=\pi/2$.  In contrast, for the vector
case we see a factor of three difference in rate between
$\triangle\phi_{jj}=0$ and $\pi$.  The flat scalar curve reflects the
lack of spin information being passed from one incoming quark current
to the other, from a $t$-channel neutral scalar Higgs boson or 4-point
$WWHH$ interaction.  As seen in Fig.~\ref{fig:scalar}, double scalar
production behaves exactly like single-scalar
production~\cite{higgs_coup}.  Thus $\Delta\phi_{jj}$ is the one
distribution we find which distinguishes the fermion and vector cases
-- and both from the scalar case.

\bigskip

\begin{figure}[ht!]
\begin{center}
\includegraphics[scale=0.4]{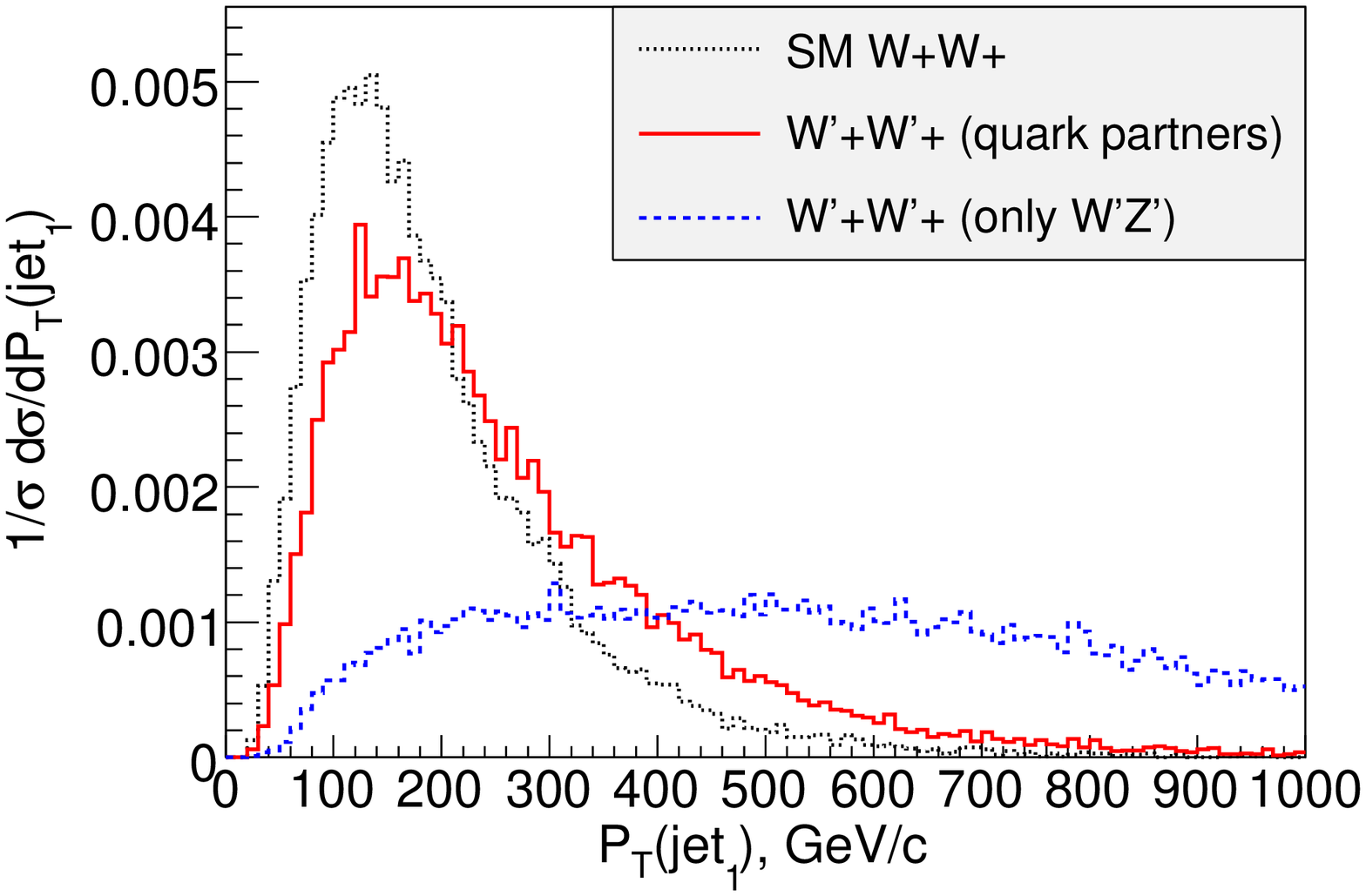}
\includegraphics[scale=0.4]{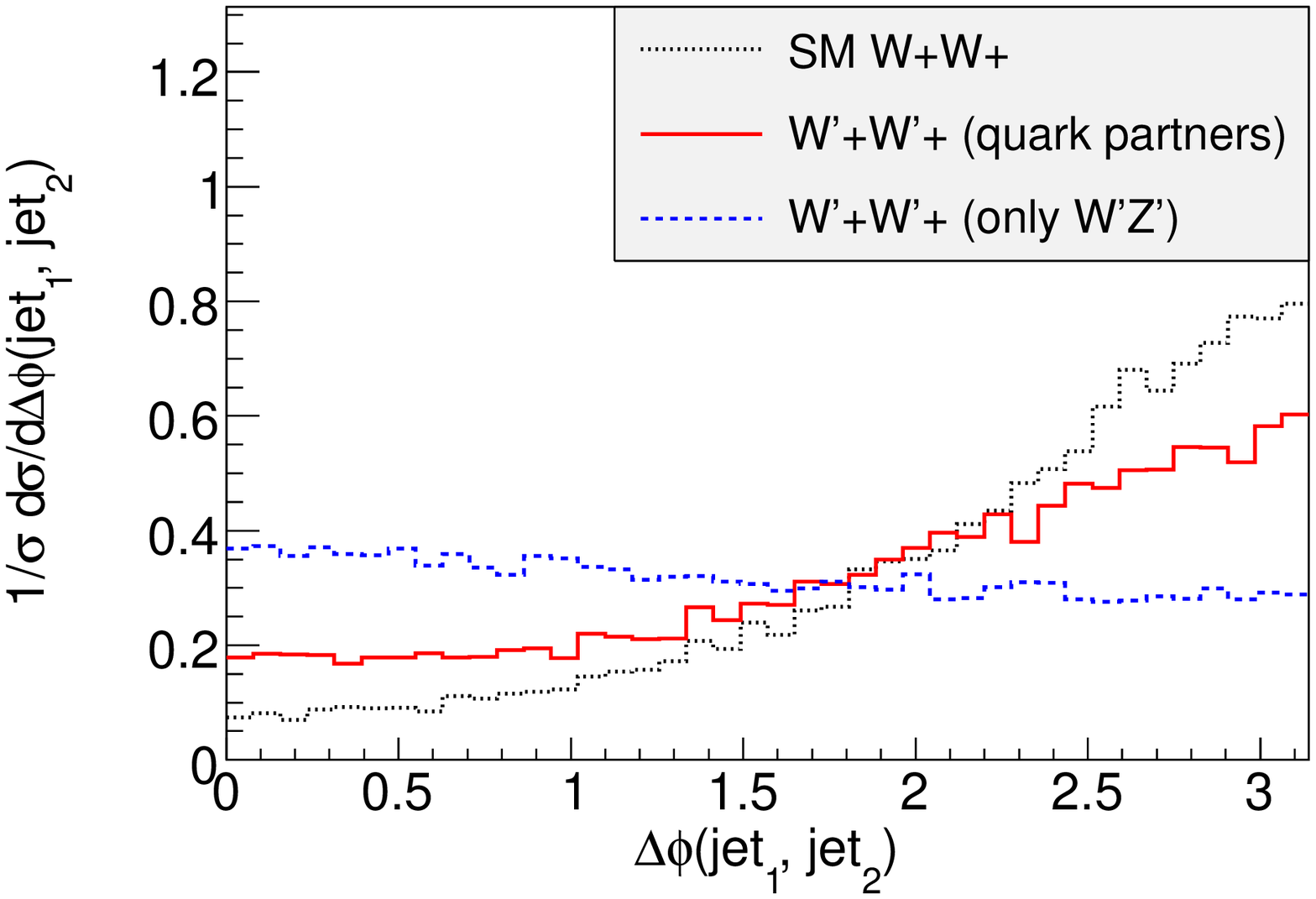}
\end{center}
\vspace{-5mm}
\caption{Comparison between Standard Model $W$ pairs, $W'$ pairs, both 
with the full set of Feynman diagrams and the $W'$ case (wrongly)
omitting all heavy quark diagrams.  Bad high-energy behavior can
clearly be seen in the transverse momentum distribution of the left
panel.  The lack of gauge cancellations also leads to incorrect
angular distributions (right).}
\label{fig:wrong}
\end{figure}

Because of the gross differences in all distributions between scalar
and higher-spin cases, identifying scalar production would be
straightforward, probably requiring less luminosity for good
statistical discrimination.  Discriminating between fermion and vector
cases is also straightforward, provided one does not cut on
$\triangle\phi_{jj}$ to reduce backgrounds.

In the general spirit of our analysis we do not use any distributions
for the charged objects themselves.  Just as in the single-Higgs
case~\cite{higgs_coup}, the truly useful information is fortunately
encoded in the forward tagging jets.

\bigskip

We should briefly comment on an aspect of models with additional
vectors, {\it e.g.} Little Higgs or universal extra dimensions.  Often
these models contain a discrete parity, like $R$ parity in
supersymmetry, to provide a dark matter particle.  We find that it is
crucial to include fermionic partners of the quarks in such cases, as
in our toy vector model.  If these are left out, gauge cancellations
between WBF and $W^\prime$ Bremsstrahlung diagrams (as occur in the
Standard Model) are spoiled, producing anomalous high transverse
momentum and invariant mass distributions for the jets and vectors.
The dramatic effect of neglecting heavy-quark diagrams on the
transverse momentum distributions and on the angular correlation can
be seen in Fig.~\ref{fig:wrong}.  We do not perform a full analysis
here, but given the bad high-energy behavior of the jet (and chargino)
transverse momentum, it seems possible that signs of unitarity
violation may appear well below the scale of assumed strong dynamics
in such models.  Additionally, a lack of gauge cancellations produces
(very) incorrect angular distributions, as seen in the right panel.


\newpage

\section{Conclusions}
\label{sec:conc}

In order to expand the capability of the LHC to explore the
electroweak sector of new physics scenarios, we have examined
electroweak production of supersymmetric same-sign charginos in weak
boson fusion.  Typical signal cross sections are known to be in the
femtobarn range, small but definitely viable for long-term
measurements, not intended to supplant discovery.  Typically, large
differences in mass scales between charginos and squarks provide for
excellent suppression of SUSY-electroweak and SUSY-QCD backgrounds to
the level of the signal, already with simple weak-boson-fusion
acceptance cuts.

Observing this signal would most importantly provide direct
confirmation that at least one neutralino is a Majorana fermion.
However, that assumes that the charged particles produced are
fermions.  We therefore showed that LHC can indeed distinguish scalar,
fermion and vector same-sign production in weak boson fusion, using
only kinematic distributions of the forward tagging jets --- most
notably the azimuthal angle between them.  That the tagging jet encode
all the necessary information to discriminate between different spin
hypotheses is fortuitous: this renders our analysis ultimately
independent of whether the heavy charged particles are quasi-stable or
decay promptly.

\bigskip

For our discrete-parity vector toy model, we encountered an
interesting aspect, that to maintain gauge invariance we need
parity-odd partners of the quarks.  This might have implications for
Little Higgs models with $T$ parity, many of which do not contain
those quark partners.  Our calculations suggest that unitarity
violation have visible effects well below the strong dynamics scale of
these models, where one would assume that new physics controls the
behavior.

\bigskip

One caveat for more general scenarios is that there will be a Standard
Model background from $W^+W^+jj$ production, which is ${\cal
  O}(100)$~fb with WBF-style cuts.  After leptonic branching ratios,
it would be within a factor of a few of the SUSY cross section.  A
detailed calculation with decays is beyond the scope of this paper,
but we expect many kinematic differences to appear between the signal
and Standard Model background, in both lepton momentum and angular
distributions.


\begin{acknowledgments}
This research was supported in part by the Swedish Research Council
(JA), the U.S. Department of Energy under grant No. DE-FG02-91ER40685
(DR).  We thank Tom Rizzo, Tim Tait and in particular Kaoru Hagiwara
for useful discussions on many aspects discussed in this paper, and
Joe Lykken for providing an incentive to speed up.  T.P. would like to
thank the DESY theory group for their hospitality where this paper was
finalized.  D.R. would like to thank the SUPA ultra-mini workshop
series for their support during his stay in Edinburgh.
\end{acknowledgments}


\end{document}